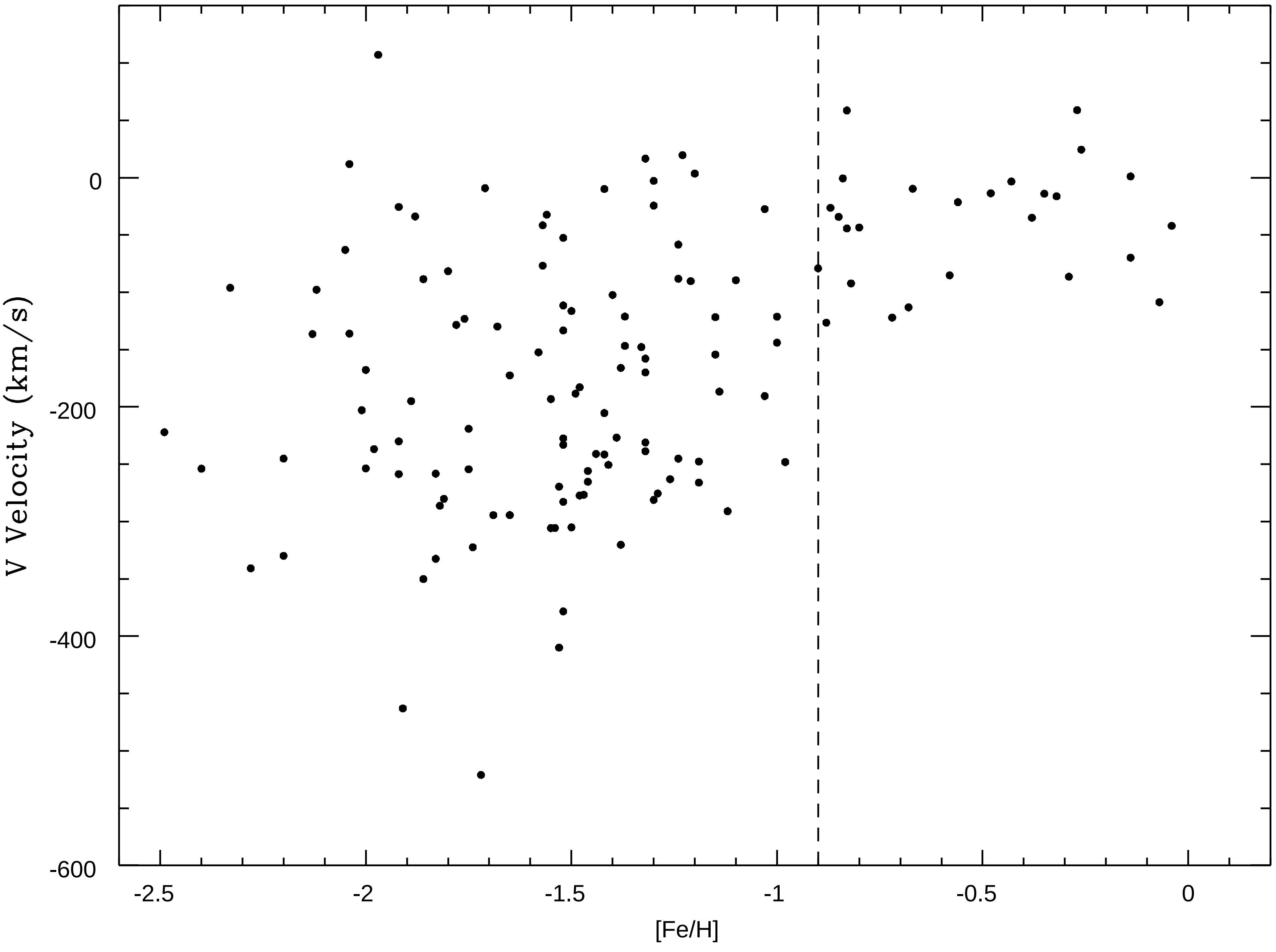

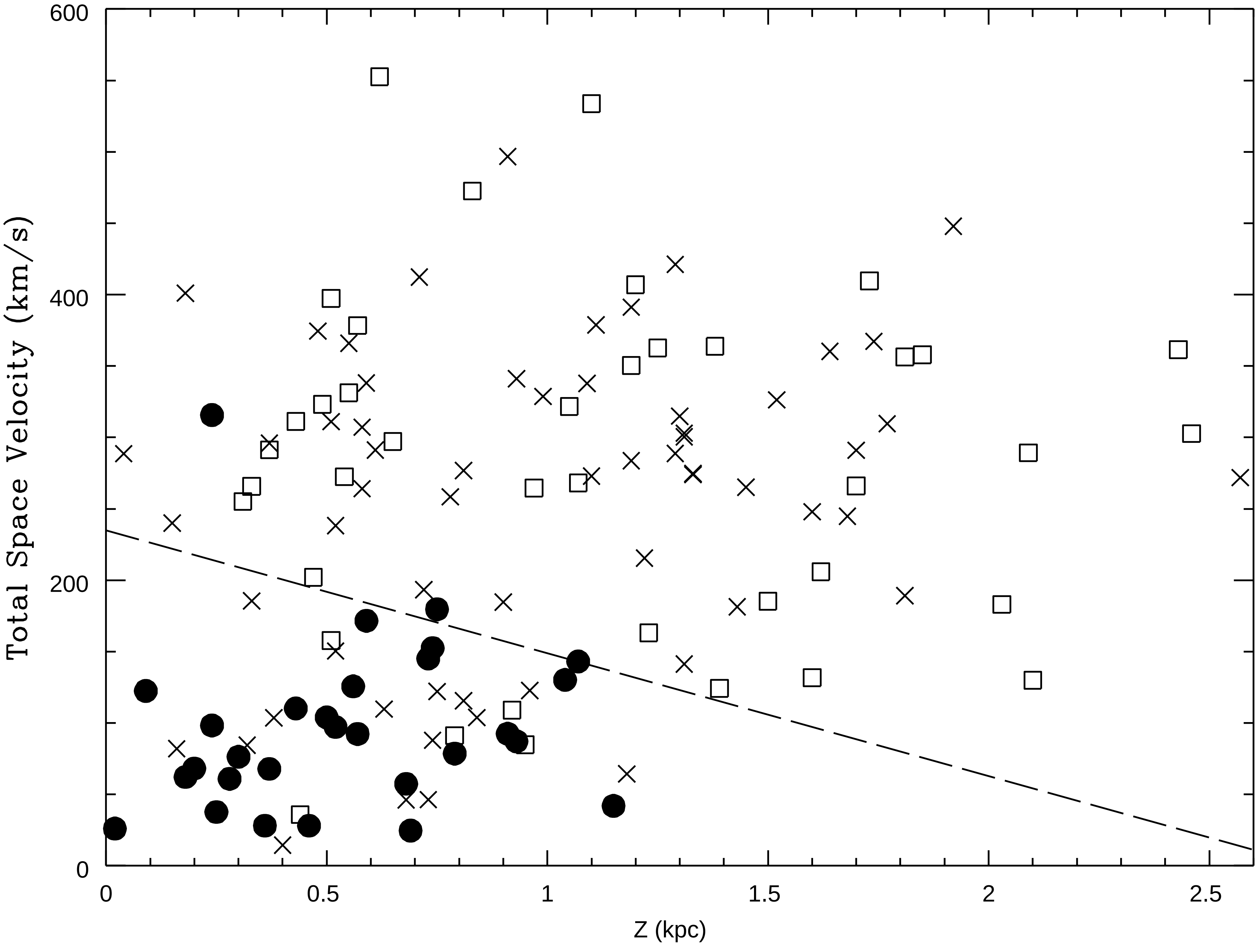

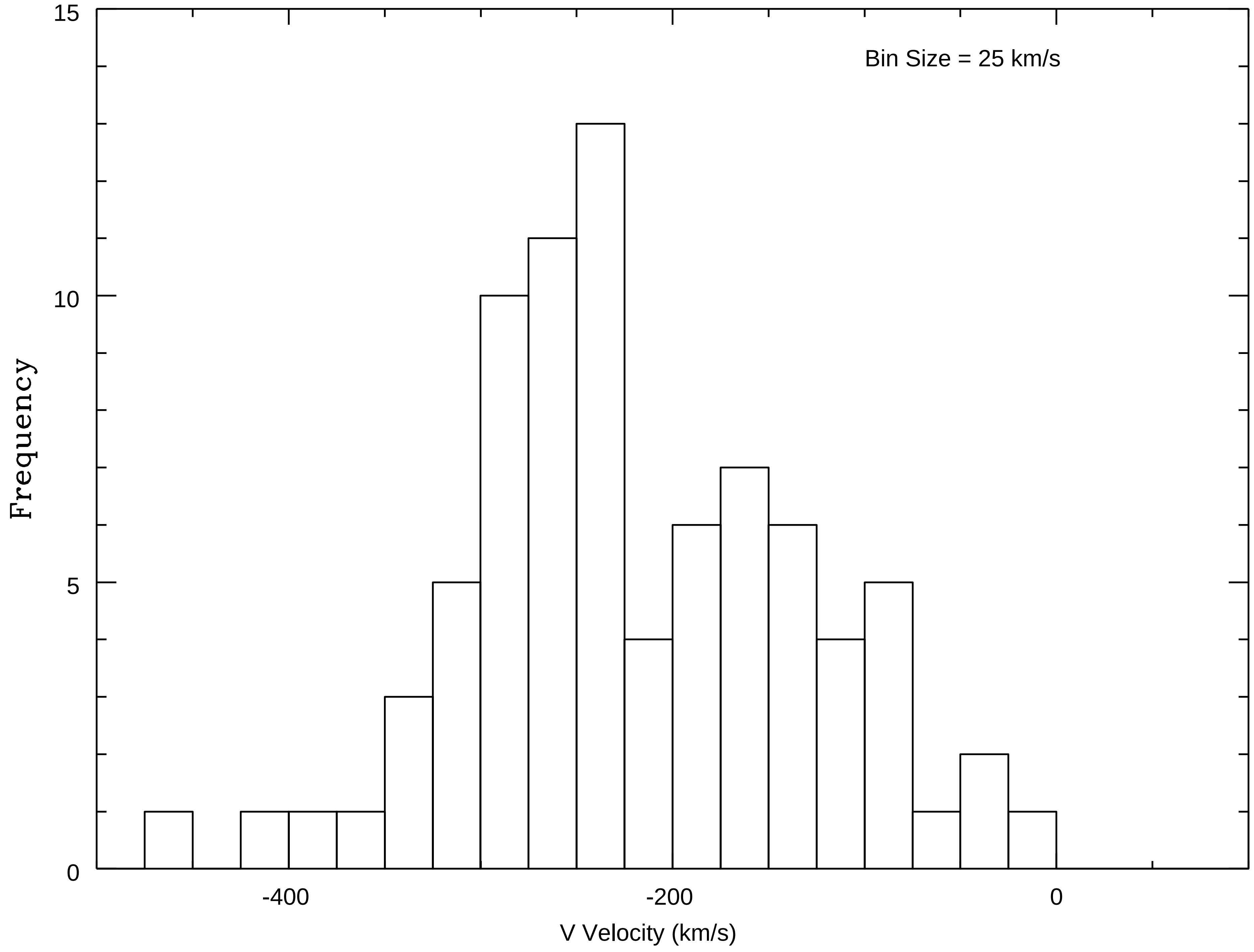

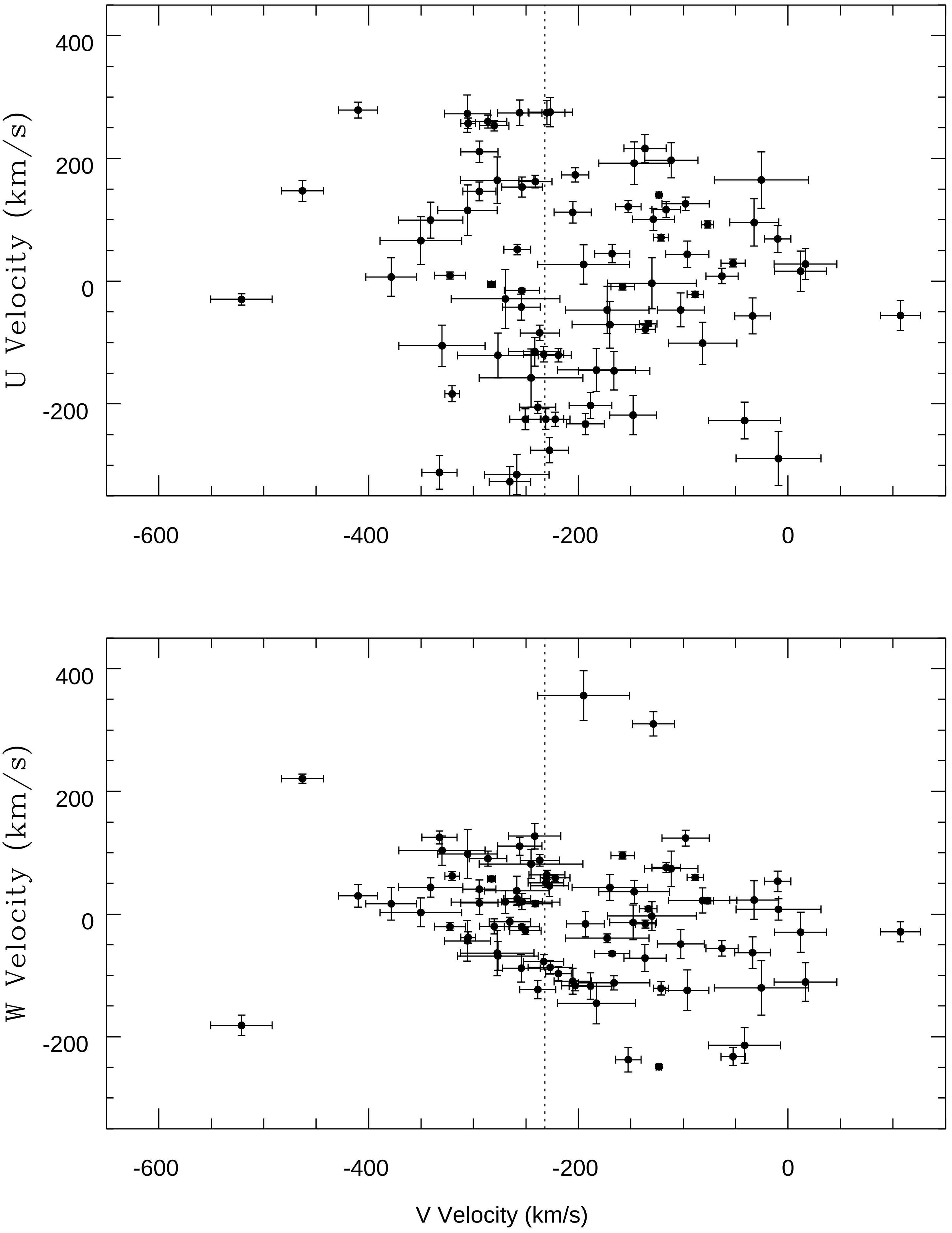

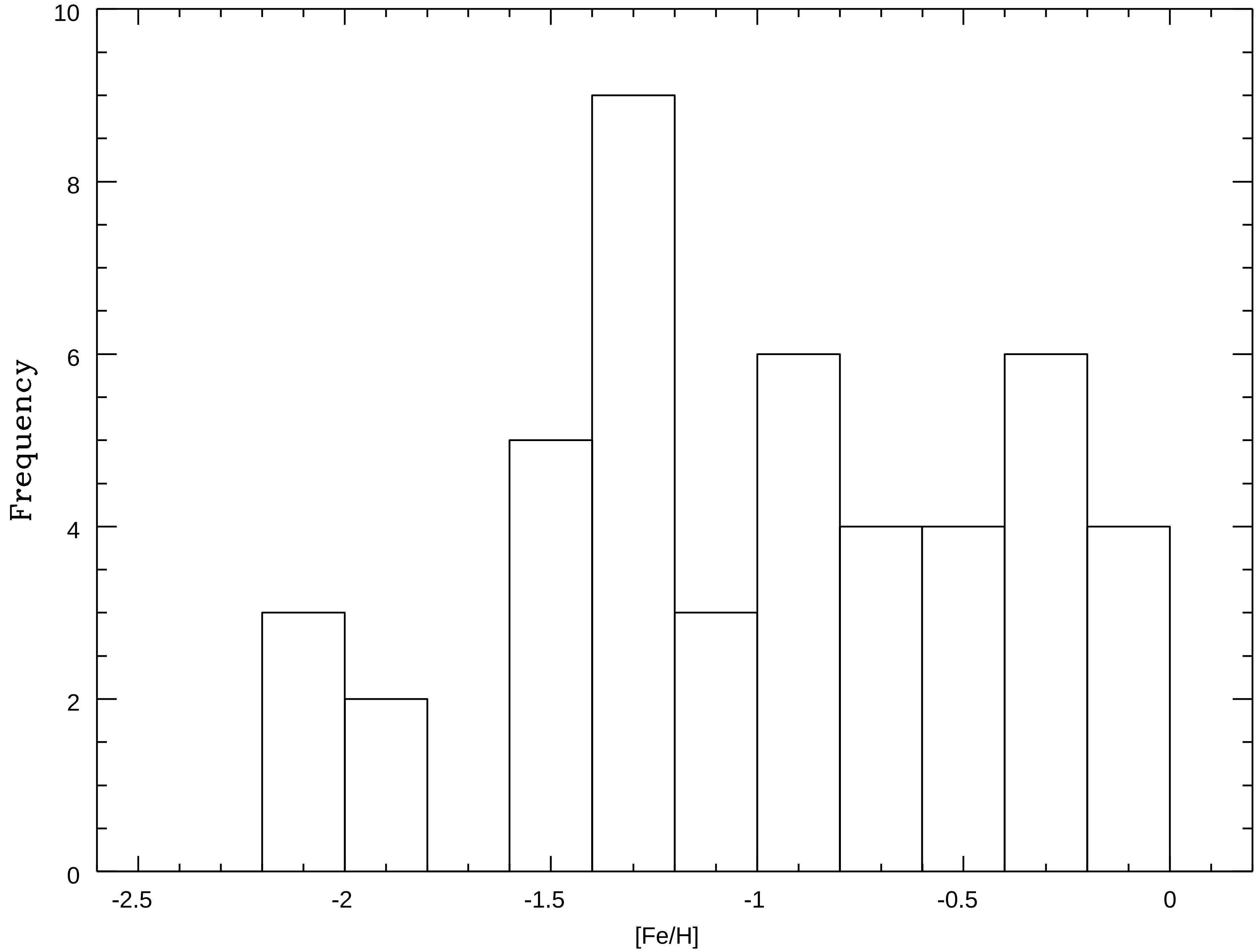

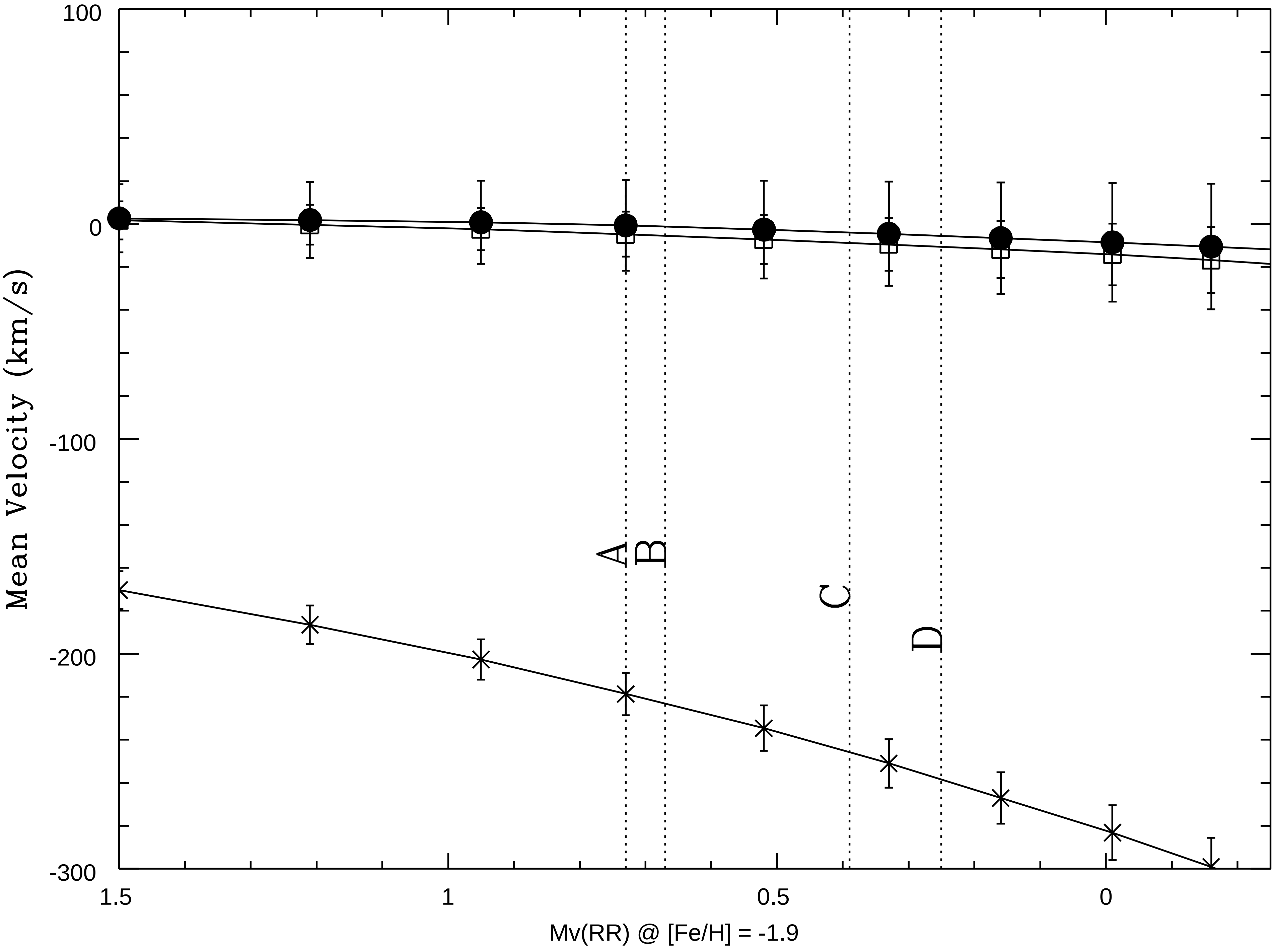

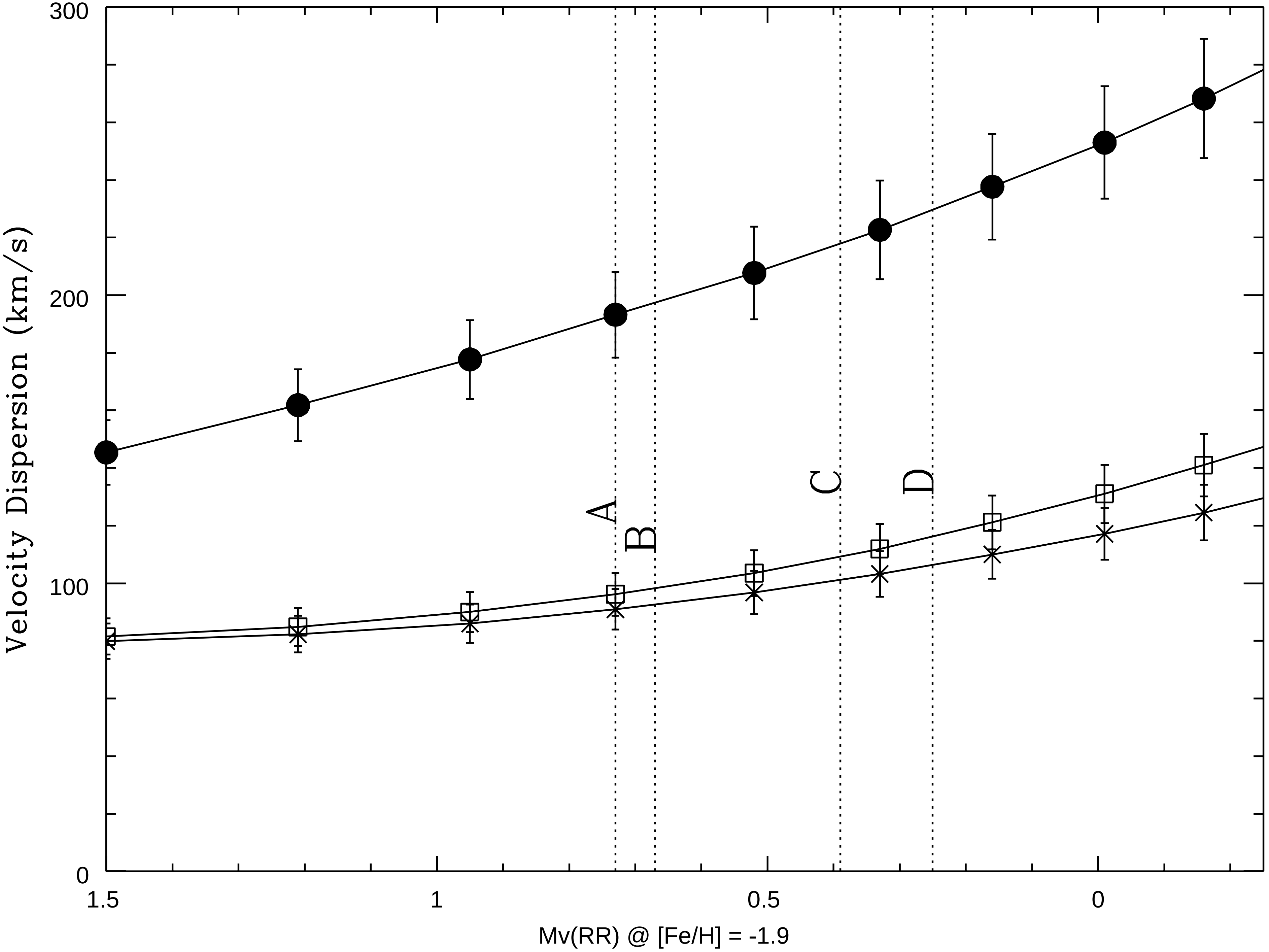

# A New Analysis of RR Lyrae Kinematics in the Solar Neighborhood

John C. Martin

Department of Astronomy,

Case Western Reserve University, Cleveland, OH 44106

Electronic mail: *martin@grendel.astr.cwru.edu*

Heather L. Morrison

Department of Astronomy and Department of Physics,

Case Western Reserve University, Cleveland, OH 44106

Electronic mail: *heather@vegemite.astr.cwru.edu*

## Abstract

Full space velocities are computed for a sample of 130 nearby RR Lyrae variables using both ground-based and Hipparcos proper motions. In many cases proper motions for the same star from multiple sources have been averaged to produce approximately a factor of two improvement in the transverse space velocity errors. In most cases, this exceeds the accuracy attained using Hipparcos proper motions alone. The velocity ellipsoids computed for halo and thick disk samples are in agreement with those reported in previous studies. A distinct sample of thin disk RR Lyraes has not been isolated but there is kinematic evidence for some thin disk contamination in our thick disk samples. Using kinematic and spatial parameters a sample of 21 stars with [Fe/H] < -1.0 and disk-like kinematics have been isolated. It is concluded from their kinematics and spatial distribution that these stars represent a sample of RR Lyraes in the metal weak tail of the thick disk which extends to [Fe/H] = -2.05. In the halo samples the distribution of V velocities is not gaussian, even when the metal weak thick disk stars are removed. Possibly related, a plot of U and W velocities as a function of V velocity for the kinematically unbiased halo sample shows some curious structure. The cause of these kinematic anomalies is not clear. In addition, systematic changes to the distance scale within the range of currently accepted values of $M_v$(RR) are shown to significantly change the calculated halo kinematics. Fainter values of $M_v$(RR), such as those obtained by statistical parallax (~ 0.60 to 0.70 at [Fe/H]=-1.9), result in local halo kinematics similar to those reported in independent studies of halo kinematics, while brighter values of $M_v$(RR), such as those obtained through recent analysis of Hipparcos subdwarf parallaxes (~ 0.30 to 0.40 at [Fe/H]=-1.9), result in a halo with retrograde rotation and significantly enlarged velocity dispersions.

## Introduction

The correlation between kinematics and metallicity gives useful information for formulating theories of galactic structure. Differences in chemistry and space velocities are crucial in defining the different populations within the Milky Way and inferring their origin. Populations of particular interest in the neighborhood of the Sun are the old thin disk, thick disk, and halo. Differences in kinematics of different populations may be subtle, so high-precision data are important.

The old thin disk is comprised of the kinematically hotter portion of the thin disk (stars older than about 1 Gyr), confined to a scale height of about 300 parsecs (Gilmore & Reid 1983). It is kinematically well mixed with an asymmetric drift of about 15 km/sec (Freeman, 1987). The metallicity distribution of the old thin disk peaks at about the solar value with approximately ±0.2 dex spread (McWilliam 1990).

The thick disk is the kinematically hottest portion of the disk of the galaxy, with a scale height of about 1.0 kpc (Gilmore & Reid 1983) and an asymmetric drift of about 40 km/s (Carney et al. 1989). Thick disk stars are the oldest stars in the disk (Edvardsson et.al. 1993) with a metallicity distribution peaking at about [Fe/H]=-0.5 (Carney et al. 1989) . There is evidence that the thick disk contains stars with metallicity as low as [Fe/H] = -1.6 or even lower (Norris et al. 1985, Morrison et al 1990, Beers and Sommer-Larsen 1995). The first two papers use samples of K giants whose metallicity was measured using the DDO photometric system. Later studies (Twarog and Antony-Twarog 1994, Ryan and Lambert 1995) showed that in the metallicity range of interest ([Fe/H] < -1.0) the DDO metallicities were systematically too low. This resulted in an over-estimate of the number of metal-weak thick disk stars, and Twarog and Anthony-Twarog concluded that "it is questionable that [the metal-weak thick disk] exists as a separate population". However, other samples, with different metallicity calibrations, have also identified metal-weak thick disk stars, and we will show in this paper that there are a small but significant number of metal-weak thick disk RR Lyraes in our sample.

The halo is characterized by a roughly spherical space distribution with close to zero net rotation (Carney & Latham 1986). Its stars are metal poor, with a peak metallicity at [Fe/H]=-1.6 (Laird et al. 1900). However, since the mid-1980's, many studies have suggested that it cannot be described by a single, smooth, and kinematically well-mixed entity. There have been several suggestions of a two-component halo, with a flattened component in the inner halo and a more spherical outer halo, including Hartwick (1987), Preston et al. (1991), Kinman et al. (1994) (who used the spatial distribution of RR Lyraes and blue horisontal-branch stars), Zinn (1993), (who used globular cluster data), Sommer-Larson and Zhen (1990), Norris (1994), and Carney et al. (1996) (who used field-star samples). It is also possible that accretions of dwarf galaxies like the Sgr dwarf (Ibata et al. 1994) make the galactic halo so complex that separation into two components is not a good description. Perhaps the halo is better thought of as resembling a “bowl of spaghetti” in phase space, as the accreted satellites slowly phase-wrap (Majewski et al. 1994, Johnston et al. 1995). We should keep in mind that investigations of halo kinematics may only be applicable to a specific place in the Galaxy (in the case of our study, the solar neighborhood) and may have velocity structure smoothed out by the velocity resolution of the study.

RR Lyraes are good tracers of these stellar populations because they are relatively bright, sample a large volume of space, have a short period of variability which make them easily identifiable, and cover a wide range of metallicities (most between -2.0<[Fe/H] < 0.0). Originally

RR Lyraes were assumed to be a fairly homogenous group mostly found in the halo. They occupy a fairly narrow region of the HR diagram where the helium burning horizontal branch crosses the instability strip between $T_{eff}$ of 6100K to 7400K (Smith 1995). The masses of RR Lyraes range from about 0.6 to 0.8 solar masses (Smith 1995), which implies ages from roughly 14 to 17 Gyrs. This mass/age bias could preclude RR Lyraes being found in the old thin disk population. The higher metallicity of the thin disk would also shift the zero age horizontal branch towards the red, out of the instability strip. For these reasons RR Lyraes are rarer in younger and more metal rich populations. Taam et al. (1976) have suggested that there is a small possibility that a higher mass star could lose enough mass while ascending the giant branch for it to land in the instability strip on the zero age horizontal branch. In this manner RR Lyraes covering a wider range of ages and metallicities could be formed. However, we lack a complete understanding of the mass loss parameters involved.

Preston (1959) was the first to make a comprehensive survey of RR Lyraes. He concluded that the RR Lyraes in his sample covered a range of metallicities and kinematics that are consistent with both the disk and halo. In Preston's magnitude-limited sample about 25% of the RR Lyraes belong to the disk and about 75% to the halo.

For nearly two decades no further *large scale* surveys of RR Lyraes were conducted. Layden (1994, 1995) made an updated survey containing a complete sample outside of the galactic plane (his survey is incomplete at galactic latitudes less than $10^{o}$) and produced improved metallicity and radial velocity data. The most important improvement over previous studies are Layden's highly accurate metallicities (see Lambert et al 1996). Layden et al. (1996; referred to here as LHHKH) added proper motions from the NPM1 (Lick Proper Motion Survey, Klemola et al., 1993) and Wan et al. (1980) to compute full space velocities in addition to adding more stars at low galactic latitudes. LHHKH concluded that RR Lyraes show two chemically and kinematically distinct populations in the solar neighborhood: the thick disk and the halo.

In LHHKH, the errors in the space velocities were dominated by proper motion errors(which are 2 to 3 times the radial velocity errors). In this work we will improve on the LHHKH space velocity errors by improving the proper motion estimates.

Errors in distance to RR Lyraes make an important contribution to errors in space velocity. With good photometry random errors are reduced to a few percent or less. Of more concern are the systematics introduced by adopting a distance scale, which vary by as much as thirty percent. Recently Feast and Catchpole (1997) and Chaboyer et al. (1998) have argued for a longer distance scale ($M_v$(RR)=0.30 at [Fe/H]=-1.9). We will discuss the effect of changes in the distance scale on our derived kinematics.

## The Database

<u>Origin and Overlap</u>

The sample of RR Lyraes we used as a basis for our database is composed of all known RR Lyrae variables north of declination -10$^o$ that are brighter than 11th magnitude as defined by Kinman (1997), who has obtained high quality light curves for the entire sample. We relied almost exclusively on metallicities, radial velocities, and distances from Layden (1994) because 89% of the Kinman sample (132 of 149 stars) are also present in that sample. Layden (1994) is an all-sky sample so 162 of Layden's stars are excluded from ours by our southern declination cut off. Because good distances, metallicities, and radial velocities already exist for most of this sample, an improvement in the proper motion data significantly reduces the errors in the computed space velocities. We have all the data needed to calculate full space velocities for 130 of the stars in the Kinman sample (128 of which appear in Layden 1994, LHHKH, or both).

Proper Motions

In order to compute full space velocities we need to have accurate distances, radial velocities, and proper motions. For this sample, average proper motion errors are around 10% while typical random distance errors are only a few percent. Proper motions are the most difficult of the three ingredients to measure because they require high precision positional data gathered over a span of at least several decades. (Space based observations now allow a similar precision in a shorter time.)

The Lick Northern Proper Motion Survey (Klemola et al. 1993; NPM) is a natural source of proper motion data for our sample because it contains many stars of astrophysical interest, including most of the RR Lyrae variables in our sample. Proper motion data was used from a number of other catalogs: the USNO Twin Astrograph Catalog (Zacharias et al. 1996; TAC), the Hipparcos Catalog (Perryman et al. 1997; HIP), the Astrographic Catalog Reference Stars (Corbin et al. 1991; ACRS), the Position and Proper Motion Catalog (Roser & Bastian 1989; PPM), and a list of proper motions of RR Lyrae stars published by the Shanghai Observatory (Wan et al. 1980l; WMJ).

The TAC is a recently published work that covers a range of apparent magnitudes slightly fainter than the ACRS or PPM with improved astrometric accuracy over both. About half of the RR Lyraes in our sample are present in the TAC. The TAC contains fewer of our RR Lyraes than the NPM because it is a magnitude limited catalog and is not compiled from a list of stars of astrophysical interest. The HIP, like the NPM, targets stars of astrophysical interest but contains fewer RR Lyraes. The HIP is of better or comparable astrometric accuracy to the NPM or TAC. The ACRS and PPM are two widely used catalogs known for good astrometric accuracy. Since neither of these catalogs contain many stars fainter than 9th magnitude, only the brightest stars in our sample of RR Lyraes are included. Data from the Shanghai Observatory catalog (WMJ) was only used in cases where there was no other source for a star's proper motion, since LHHKH found that the error estimates for the WMJ proper motions are unreliable.

We employed an average weighted by the inverse variance for all the stars which had proper motions independently determined in two or more catalogs. This scheme, in addition to reducing the errors, has the advantage of reducing the influence of any small systematic errors in the individual catalogs. Proper motions for 80 of the 130 stars in the sample were improved in this manner (see Table 1). The proper motion data for the stars in our sample are given in Table 2a.

It should be noted that the proper motions in all the catalogs except the NPM and HIP are on the FK5 J2000 system. The HIP is on the International Celestial Reference System (ICRS) which has replaced the FK5 system. The ICRS is consistent with the FK5 J2000 coordinate system so any differences are not significant (Arias et al., 1995). The NPM proper motions are based on an "absolute" frame that appears to have no significant formal errors with respect to the HIP proper motions on the ICRS system (less than 1 milli-arcsecond per century) (van Leewen et al. 1997). Thus the all proper motions in this study have been treated as if they are on the same astrometric reference system.

Distances, Metallicities, and Radial Velocities

We used distances from Layden (1994). These employed a value of $M_v$(RR) of 0.73 at [Fe/H] = -1.90. A small improvement in the absolute magnitude calibration used to determine the distance for RR Lyraes was published in LHHKH ($M_v$(RR)=0.67 at [Fe/H]=-1.90). When this correction is applied to the Layden (1994) distances, it results in a systemic shortening of the distances by 2.3%. This factor has no significant effect when compared to the random errors quoted for the distances, which are on the order of 10%. A more dramatic change in the RR Lyrae luminosity calibration ($M_v$(RR)=0.25 at [Fe/H]=-1.90), proposed by the results from Hipparcos parallaxes and a revision of the distance to the LMC (Feast & Catchpole 1997), lengthens the distance scale by 25%. The effect of this on our results will be discussed in the section "Changes to the Distance Scale."

The metallicities and radial velocities for our sample were taken from Layden (1994) and LHHKH. The ΔS relation used by Layden (1994) and LHHKH to measure the metallicities of the RR Lyraes in his sample is calibrated on the Zinn-West (1984) abundance scale. The ΔS to [Fe/H] relation has since been re-examined by Lambert et al. (1996). They showed that to a high degree of accuracy the Layden (1994) [Fe/H] values agree with values derived from high S/N, high resolution spectra of Fe II lines.

Eight of the stars in the sample did not have distances or radial velocities in Layden (1994) (see Table 2b). Six of these are in LHHKH, though no errors are give for the [Fe/H] or distance values. For two of these stars radial velocity data was obtained from the Hipparcos Input Catalog (HIC; Turon et al. 1992) and [Fe/H] values were computed using ΔS values from Preston (1959) and the ΔS to [Fe/H] relation from Layden (1994). Photometry was obtained from Kinman (1997) to calculate the distances to all eight stars. The photometry included mean apparent V magnitude and (B-V) colors at minimum light. The (B-V) colors at minimum light were used to obtain interstellar extinction factors, following Blanco (1992) and assuming a reddening coefficient (R) of 3.20. The mean absolute V magnitude for each star is computed using the method of LHHKH. The extinction, mean absolute V magnitude, and the mean apparent V magnitude are then combined to calculate a distance. Errors in this distance are computed by a standard Monte-Carlo error simulation. The random errors for the distances computed from the Kinman photometry averaged just under 2%, compared to 8% for the LHHKH distances.

## Numerical Methods

Coordinate Transforms

The equatorial coordinates in the database were converted into galactic coordinates using standard transformations. The galactic coordinates were in turn used to obtain galacto-centric distance R and height above the plane Z. For these last calculations the position of the Sun was assumed to be 8 kpc from the galactic center and in the plane of the Galaxy (Z=0).

Space Velocities

The U, V, and W space velocities are computed from the distance, radial velocity, and proper motion as a function of celestial equatorial coordinates using the method of Eggen (1961).[1] Errors are estimated from the errors in the position, proper motion, radial velocity and distance by a Monte-Carlo method that simulates the quoted errors in each coordinate as a one sigma random variation of a gaussian distribution.

The mean U, V, and W velocity and velocity dispersions for a sample are calculated using a trimmed mean and sigma routine (Morrison et al. 1990). In this case, ten percent of the most extreme values are excluded from the calculations, making the results less sensitive to outliers.

## Population Analysis

Defining Galactic Populations

There are four broad parameters which can be used to define distinct populations of stars in our Galaxy: position, chemical composition, kinematics, and age. The simplest way to split the disk and halo populations is to divide them chemically. Stars with [Fe/H] < -1.0 are predominantly of the halo population and stars with [Fe/H] > -1.0 are mostly members of the disk population. However, this method ignores the overlap in [Fe/H] between the two populations. A more sophisticated method will attempt to sort out the kinematic, chemical, and spatial overlaps between populations. Age, determined by fitting stars to calibrated isochrones, can sometimes be used to distinguish populations. However, it is difficult to measure ages for RR Lyraes to any accuracy so we will not discuss age any further. The overlaps between populations are of particular interest because they provide insight about the formation histories.

It is important to keep in mind biases that may arise in defining samples. A sample defined by a property such as metallicity or kinematics will yield results biased with respect to that property. As an example, it is necessary to use a kinematic and spatial definition of a population to study the metal weak thick disk so as to not bias the disk population against metal weak stars.

Kinematically Unbiased Samples

Initially, disk and halo populations were separated by metallicity to yield a kinematically unbiased sample for analysis and then by kinematics and position to yield a chemically unbiased sample. Although these methods of separation do not introduce kinematic or chemical bias in each case, we should keep in mind the mass and age biases inherent to a sample of RR Lyraes.

The most dramatic difference between disk and halo populations is their rate of rotation (V velocity). Figure 1 shows the V velocity plotted as a function of [Fe/H] for our sample. A clear change in the distribution is seen at [Fe/H] = -0.9. At this point the V velocity dispersion increases and the mean V velocity changes drastically, due to the onset of the halo population. The DISK1 sample is accordingly defined as all stars with [Fe/H] > -0.9. The halo population is composed of the majority of the remaining stars. To eliminate the metallicity overlap of the disk on the halo distribution the HALO1 sample is defined as those stars with [Fe/H]<-1.3. This boundary is chosen conservatively for two reasons; first to eliminate as many low metallicity thick disk stars as possible and second to account for measurement errors in [Fe/H] that may blur the boundary between populations.

The results of the kinematic analysis of the HALO1 and DISK1 samples appear in Table 3. The velocity dispersions of the HALO1 sample are consistent with samples of the local halo using a variety of tracers including RR Lyraes (LHHKH; Layden 1995; Chiba & Yoshi 1998) , red giants (Morrison et al. 1990; Chiba & Yoshi 1998), a compilation of metal poor stars from a variety of sources without kinematic bias (Beers & Sommer-Larsen 1995), high proper motion subdwarfs (Carney et al. 1996), and a synthesis of results from many different types of tracers (Norris 1986) (see Table 4a). The Chiba and Yoshi (1998) sample is a combination of RR Lyrae and red giant stars. We prefer to focus on their results for RR Lyraes since the uncertainties in the metallicities of their red giants translate into distance errors which exceed those for the RR Lyraes, significantly enlarging the velocity errors from proper motions. The mean V velocity of the HALO1 sample (-197 ± 12 km/s) is consistent with a slightly prograde halo with $V_{rot}$=35±12 km/s(taking the LSR rotation to be 220 km/s and the Sun's velocity to be +12 km/s). The U velocity dispersion (180 ± 14 km/s) is slightly larger than other estimates but agrees within one sigma with other studies.

The DISK1 sample has velocity dispersions similar to several published thick disk samples, including RR Lyraes, F subdwarfs, and proper motion selected samples (LHHKH; Layden, 1995; Edvardsson et al., 1993; Beers & Sommer-Larsen, 1995; see Table 4b). The DISK1 sample has an asymmetric drift of +41 ± 11 km/s, also consistent with the other thick disk samples. However the mean W velocity of the DISK1 sample is larger than we should expect. The contribution of solar motion to our mean W is only -7 km/s (Mihalas & Binney 1981) and the mean for our sample is -29 ± 6 km/s.

The reason for this discrepancy in the mean W velocity for the DISK1 sample is uncertain.

No other study shows such a large negative mean W velocity. LHHKH found a more negative than normal mean W velocity in their sample of RR Lyraes (-16 ± 6 km/s) which has a two sigma overlap with our value. The plot of W velocities of the RR Lyrae sample from Chiba and Yoshi (1998) also shows the same lack of metal rich RR Lyraes with positive W velocities. Although our sample is not an all sky sample like LHHKH or Chiba and Yoshi (1998) they find the same effect to a lesser extent, suggesting that spatial sampling is not the cause. Also, we have been unable to identify "moving groups" in the DISK1 sample that may be biasing our results.

Since many of the stars in the thick disk (DISK1 sample) are observed at low galactic latitudes and our sample is more complete here than LHHKH's, the transverse component of the velocity dominates the sample's calculated mean W velocity. Thus possible errors in proper motions and distance need to be considered carefully. The proper motions for the stars in the DISK1 sample came from almost every proper motion source in the database, eliminating the possibility of a systematic effect from a single catalog. Could this drift be in the reference frames of the catalogs? This possibility seems very unlikely since other proper motion surveys utilizing the same coordinate systems have not obtained similar results.

Changing the RR Lyrae distance scale does not resolve the problem. Adopting the extreme value of $M_v(RR) = 2.23$ at $[Fe/H] = -1.9$ results in a mean W velocity of -16±5 km/s for the DISK1 sample. However in this case the mean V velocity becomes -23±7 km/s and the velocity dispersions are $(\sigma_U, \sigma_V, \sigma_W)$=(42±6 km/s, 36±5 km/s, 23±3 km/s), values typical of the thin disk, not the thick disk.

Thus we have been unable to identify the reason for the non-zero mean W velocity. A larger sample may help identify the factor influencing our result.

The Thin Disk

The DISK1 sample appears to be only representative of the thick disk population, not the thin disk population. If RR Lyraes do exist in the thin disk then they most likely exist in small numbers only and it would be difficult to separate them out of our small sample of 26 stars.

The W velocity dispersions calculated for the thick disk using RR Lyraes (See Table 4b, LHHKH, Layden 1995) are smaller than W dispersions calculated using other tracers (Beers & Sommer-Larsen 1995, Edvardsson et al. 1993). This could be a small thin disk contamination of the RR Lyrae thick disk samples. However, there is no significant difference in the asymmetric drift of the thick disk in RR Lyraes samples as would be expected with significant thin disk contamination. Since the error in the Z velocity dispersion is smaller we might expect it to be more sensitive to a small amount of thin disk contamination. Because of the size of the sample, our results are inconclusive as to the existence of thin disk RR Lyraes. When new data are available for the entire Kinman sample (Kinman 1997, Morrison et al. 1998) this situation may improve, as of the 19 stars in the Kinman sample which are not in this work, 11 have galactic latitudes less than 30$^{o}$, so are likely to be disk stars.

Chemically Unbiased Samples

A plot of total space velocity (relative to the Sun; $V_{tot}^2 = U^2 + V^2 + W^2$) as a function of Z (height above the plane of the disk) shows that almost all of the stars classified as thick disk stars in the DISK1 sample have low space velocities and small distances from the galactic plane (Figure 2). To kinematically and spatially separate the thick disk and halo populations a line was drawn: $V_{tot}$ (km/s) = 235 - 86 * Z (kpc) (dotted in Figure 2)[2]. Those stars in the region above this line were placed in the HALO2 sample and those below the line in the DISK2 sample. A slanted line is used to separate the samples because the sum of a single star's potential energy (represented by Z) and kinetic energy (represented by the total space velocity) should fall in different ranges for each of the two populations. Thus a star in the disk could have a large total space velocity if its distance from the galactic plane was proportionally smaller. Note in Figure 2 that the region of small $V_{tot}$ and large Z is unpopulated. This is because it is unlikely that a star far from the galactic plane will be moving in a circular orbit like the Sun.

The Halo

The HALO2 kinematics are similar to those of the HALO1 sample except on two points. First HALO2 has a somewhat larger U velocity dispersion (193 ± 15 km/s) than the HALO1 sample (180±14 km/s). Both are larger than is typical of a local halo sample (Beers & Sommer-Larsen 1995). HALO2 also shows a halo with no net rotation (<V> = -219 ± 10 km/s; $V_{rot}$=+13±10 km/s). It is possible that this is the more correct result because some of the lowest metallicity stars with disk like kinematics remained in the HALO1 sample and would be responsible for the resulting slight prograde rotation. The DISK2 population contains stars with disk-like velocities and metallicities as low as [Fe/H]=-2.0, with 12 stars having [Fe/H] < -1.3.

The kinematics of our HALO2 sample are consistent with those calculated by Carney et al. (1996) for a "low" halo sample selected by orbital eccentricity (a method which should exclude most metal weak thick disk stars from that sample). However, the "low" halo sample selected by Carney et al. (1996) by metallicity shows a stronger prograde rotation, having kinematics more consistent with our HALO1 sample (see Table 4a).

A histogram of the V velocities of the stars in the HALO2 sample does not have a gaussian shape (see Figure 3). It appears bimodal, with the division at $V_{rot}$~0 (V= -232 km/s). Varying the histogram bin size and location does not significantly alter this distribution. Plots of both U and W velocity against V velocity for the HALO1sample (Fig. 4) show some curious structure: the stars with retrograde orbits have a lower W velocity dispersion than the prograde stars, while the reversee is true for the U velocity. Is there a real difference between the prograde and retrograde halo stars, perhaps suggesting a different origin? Both groups have a similar [Fe/H] distribution, and both are similarly distributed on the sky. Also, different values of Mv(RR) do not substantially change this result.

What might be the cause of the differences seen in Fig. 4? The clumping in W velocity suggests the possibility of moving groups (although we would expect to see a similar amount of clumping in U velocity). Johnston, Spergel, and Hernquist (1995) showed that a tidally disrupted group of stars in the galactic halo should spread out along the orbit of the original group, maintaining a small velocity dispersion along the axis perpendicular to the orbital motion. We were unable to subdivide any portion of our halo samples into moving groups with these unique kinematic signatures. However, a portion of the halo consisting of many of these tidally disrupted groups may have a kinematic signature that differs from the gaussian velocity distributions expected for the halo. A more extensive sample is needed to investigate this further.

The Disk

The [Fe/H] distribution of the DISK2 sample (see Figure 5) includes a significant number of metal weak stars ([Fe/H]<-1). The DISK2 sample is broken into two additional samples which are also analyzed in Table 3. The DISK2A sample includes the DISK2 stars with [Fe/H] less than -1.0 and the DISK2B sample includes all of DISK2 with [Fe/H] greater than -1.0.

The kinematics of the DISK2B sample are almost identical to those of the DISK1 sample, which is to be expected since they contain almost all the same stars. The average V velocity and the velocity dispersions of the DISK2A sample not significantly different from those calculated for the DISK1 and DISK2B sample or other thick disk samples (see Table 4b). We believe the slightly larger values are due to a small amount of halo contamination in the DISK2A sample.

Dropping the four stars with V velocities less than -200 km/s from the DISK2A sample changes the average V velocity to -41 km/s and decreases each of the velocity dispersions by about 10 km/s. The resulting kinematics are a closer match to the other thick disk samples. The DISK2A sample is clearly taken from the thick disk population but contains stars with more halo-like metallicities.

The Metal Weak Thick Disk

We propose that the DISK2A sample is taken from the metal weak thick disk first identified by Norris et al. (1985). The mean velocities and velocity dispersions of our DISK2A sample are similar to those calculated for the metal weak thick disk by Morrison et al. (1990) (see Table 5).

We can compare the number of RR Lyraes in the metal weak thick disk to the number in the halo because they cover the same abundance range. Using our DISK2A and HALO2 samples, $N(RR)_{MWTD}/N(RR)_{HALO}$ is 0.26 ± 0.06. Layden (1995) estimated the number of kinematically disk-like RR Lyraes with $-1.6 \leq [Fe/H] < -1.0$ in a region of space within 1 kpc of the plane. Our results are within the range which Layden estimated for the ratio of thick disk to halo stars with those parameters. Chiba and Yoshi (1998) find $N(RR)_{MWTD}/N(RR)_{HALO}$ is about 0.3. This is also consistent with our result. These ratios are significantly smaller than $N_{MWTD}/N_{HALO}$ of 0.50 for G and K giants proposed by Morrison et al. (1990), because the DDO metallicity calibration that they used made some moderately metal-poor thick disk stars have [Fe/H] < -1.0 (see Twarog and Anthony-Twarog 1994).

The proportion of thick disk stars with [Fe/H] <-1 can be figured using the approximate relative numbers of thick disk and halo stars in the solar neighborhood. Morrison (1993) found $N_{Halo}/N_{TD}=1/50$. Combining this with our ratio of metal weak thick disk (MWTD) RR Lyraes to halo RR Lyraes, we obtain $N_{MWTD}/N_{TD}$=0.005±0.001. Thus, though we have shown that there are metal-poor stars in the thick disk, they form only a very small tail of its metallicity distribution.

Beers and Sommer-Larsen (1995) published a list of possible metal weak thick disk stars selected from their data by taking stars with [Fe/H] < -1.0 and radial velocities indicating a rotational velocity of less than 100 km/s. Of those stars, four are also present in our study (XX And, EZ Lyr, SW Aqr, and VV Peg). We have identified XX And and SW Aqr as belonging to the halo and EZ Lyr and VV Peg as being members of the metal weak thick disk. Since Beers and Sommer-Larsen had only radial velocities and no proper motions for the stars in their study, they had less kinematical information for each star. Using full space velocities, we have been able to more finely separate our halo and metal weak thick disk samples than Beers and Sommer-Larsen. Because of their selection criterion and limited kinematic data, it is possible that they have also mis-identified some metal weak thick stars as halo stars. For these reasons our metal weak thick disk sample represents a more complete sample with less halo contamination. Beers and Sommer-Larsen also found an extended tail to the distribution, stars with [Fe/H] < -1.6 and disk-like kinematics. Our DISK2A sample contains five stars with [Fe/H] < -1.6 with the lowest being [Fe/H]= -2.05, a significant detection of the extended metal weak tail, in agreement with their results.

The presence of the metal weak thick disk among the stars in our study also supports the previous assertion that the HALO2 sample is a better gauge of halo kinematics than the HALO1 sample. There are 11 members of the DISK2A metal weak thick disk sample that have [Fe/H] less than -1.3 and would have contributed to a slightly prograde rotation of the HALO1 sample. Removing the contamination of the metal weak thick disk we obtain the HALO2 sample which shows a non-rotating local halo. Morrison et al.(1990) also removed the metal weak members of the thick disk from their halo sample, but in that case it resulted in a halo sample with a somewhat more prograde $V_{rot}$ (25 km/s vs. 13 km/s). Having full space velocities has allowed us to more effectively remove the metal weak thick disk stars from our HALO2 sample.

Table 4b shows that the metal weak thick disk (DISK2A) has kinematics consistent with samples of metal enriched thick disk stars. Figure 2 shows that the metal poor stars ([Fe/H]<1.0) with disk-like kinematics are kinematically and spatially well mixed with the metal rich stars ([Fe/H]>-1.0). This leads us to the conclude that the metal weak thick disk is the metal weak tail of the thick disk and not a distinct population by itself and also that these stars are not a moving group in the halo.

<u>Changes to the Distance Scale</u>

Recent studies using Hipparcos data have suggested that a change is needed in the RR Lyrae distance scale. Feast and Catchpole (1997) concluded from a re-calibration of the Cepheid distance scale and application of their findings to RR Lyraes in the LMC that RR Lyraes are 0.48 magnitudes brighter than previously thought. Chaboyer et al. (1998) have used Hipparcos parallaxes to sub-dwarfs and main sequence fitting to re-examine the distances to globular clusters. They combined the values for $M_v$(RR) obtained from the new cluster distances with other $M_v$(RR) determinations to arrive at a value for $M_v$(RR) of 0.39 at [Fe/H ]=-1.9. They point out that with their new RR Lyrae distance scale, ages derived from globular cluster color magnitude diagram fits and from the Hubble constant are no longer discrepant with standard ($\Lambda$ = 0) cosmological models. We have investigated the effect of the revised RR Lyrae distance scale on the kinematics of our RR Lyrae field star sample. Such lengthening of the distance scale causes no significant changes to the kinematics we derive for the *disk* populations because the distances to these stars are smaller so changing the distance scale has a less pronounced effect on their calculated transverse velocities. However, the mean velocities and dispersions in the halo populations are altered significantly.

Figures 6 and 7 show the change in mean velocity and velocity dispersion as a function of $M_v$(RR) for the HALO2 sample. $M_v$(RR) of 0.73 is used by Layden (1995) and our study , $M_v$(RR) of 0.25 corresponds to the Feast and Catchpole value, and Mv(RR) of 0.39 is the Chaboyer et al. value. Note that the mean rotational velocity (V) changes significantly as a function of $M_v$(RR). A change of as little as 0.20 magnitudes in either direction changes $V_{rot}$ of the halo from prograde to retrograde. A similar change in distance scale also makes a significant change in the U velocity dispersion. Note that the rate of change in U dispersion as a function of distance scale is significantly different from the rates of change in V and W dispersion. These rates show that a change in distance scale has the effect of stretching or compressing the velocity ellipsoid.

Ryan (1992) pointed out that if a 16% longer distance scale is adopted for the UBV spectroscopic parallax technique used by Majewski (1992) that the retrograde rotation of the halo found in his work is reduced from $V_{rot}$= -55 km/s to $V_{rot}$= -9 km/s. Similarly, our data exhibits the same retrograde halo rotation as Majewski (1992) if we apply a lengthening to our distance scale of a factor of 20%-30%. Majewski (1992) reported that his measurement of retrograde rotation in the halo could be a product of a systematic error in the distance scale but dismissed this possibility after analysis of possible errors. Carney et al. (1996) reported local or “low” halo kinematics similar to those we have calculated for our halo samples and that the kinematics of the distant or “high” halo are consistent with those found by Majewski (1992). This would imply that the portions of the halo sampled by Majewski (1992) are dominated by a population or populations with kinematic properties different from those of the local halo. In this case we would not expect to find a strong retrograde rotation in our halo samples.

A change in the distance scale affects the computed velocity dispersions as well as the mean rotational velocity. In the case of our data, the velocity dispersions for the halo computed using $M_v$(RR) brighter than 0.40 are much larger than any other dispersions reported in the literature for other types of tracers (see Table 4a). A slight shortening of the distance scale to ($M_v$(RR) ~ 1.0 at [Fe/H]=-1.9) would actually improve the agreement of our velocity dispersion values with those previously published by decreasing the U velocity dispersion to a smaller, more frequently quoted value.

It is important to note a discrepancy between values of $M_v$(RR) arrived at for cluster and field RR Lyraes. This was first noted by Chaboyer et al. who left the results from the analysis of field stars out of their analysis $M_v$(RR). This disagreement is troubling because the kinematics of the halo are significantly changed by adopting different values of $M_v$(RR) within the current acceptable range of values. The brighter values of $M_v$(RR), adopted from analysis of cluster RR Lyraes, indicate a halo with larger velocity dispersions and retrograde rotation, while the fainter values of $M_v$(RR), arrived at from field RR Lyraes, indicate kinematics similar to those appearing in other independent kinematic analyses of the halo. Catelan (1998) has found no difference between the period-temperature distributions of field and cluster RR Lyraes, ruling out the possibility of two groups differing in physical properties. It seems likely that systematic errors may be responsible for this discrepancy rather than a fundamental physical difference between cluster and field RR Lyraes.

**Summary and Conclusions**

The results of our kinematic analysis of disk and halo samples agree in general with other published results (Table 4a and IIIb). It is our belief that the HALO2 sample (defined as stars with total space velocities greater than 235 km/s - 86 * Z, where Z is the height above the galactic plane in kpc), despite being kinematically biased, better represents the true kinematics of the halo since fewer thick disk stars with small [Fe/H] are present in this sample than the HALO1 sample.

The computed W velocity dispersion for the DISK1 and DISK2 samples are smaller than normally noted for the thick disk. (See Table 4b) Thus some thin disk stars may have contaminated our thick disk sample.

The HALO1 sample has curious kinematic structure visible in plots of U and W velocity plotted against V velocity (Figure 4). Also, a histogram of V velocities in the HALO2 sample (Figure 3) reveals a non-gaussian profile. A more extensive sample is necessary to determine the nature of these kinematic distributions and what they may tell us about the structure and evolution of the local halo.

The spatial and kinematic parameters used to separate the HALO2 and DISK2 samples allowed us to detect an extended metal weak tail in the DISK2 distribution. We believe this tail (DISK2A) is a representative sample of the metal weak thick disk of Norris et al.(1985). The kinematic parameters we derive for the DISK2A sample are in agreement with those derived by Morrison et al.(1990) and consistent with those calculated for the more metal enriched thick disk. We find a significantly smaller proportion of metal weak thick disk stars ($N(RR)_{MWTD}/N(RR)_{HALO} = 0.26 \pm 0.06$) than Morrison et al. (1990) and that the distribution of stars in the metal weak component of the thick disk extends to metallicities at least as low as [Fe/H]=-2.0, in agreement with Beers & Sommer-Larson (1995).

With respect to the distance scale we found that a change in $M_v(RR)$ has no significant effect on the calculated kinematics of our *disk* samples. However, a shift of as little as 0.10 mag. in $M_v(RR)$ has a significant effect on the mean rotational velocity and the velocity dispersions of the halo. If we were to adopt the distance scales of Feast and Catchpole (1997) or Chaboyer et al. (1998) this would significantly enlarge the calculated U, V, and W velocity dispersions well beyond normally accepted values. Accepting this distance scale would also result in a calculated retrograde rotation of our local halo samples comparable to that detected by Majewski (1992) for the distant halo.

## Acknowledgments

We would like to thank Dr T.D. Kinman for starting the survey that inspired this work, and Anne Fry for helpful comments on an earlier draft of this paper. We would also like to thank Robert Hanson and Arnold Klemola for their help in understanding the details of the Lick NPM1 proper motion reference system. We also thank the anonymous referee for their comments, which helped enhance the quality of this paper. This work was partially supported by NSF grant AST-9624542 to HLM, and the Jason J. Nassau Scholarship Fund to JCM.

**Footnotes**

1 U is defined as the radial motion with respect to the Sun with motion towards the galactic anti-center being positive. V is defined as the rotational motion with respect to the Sun with motion in the direction of galactic rotation being positive. W is defined as motion in the Z direction with respect to the plane of the galaxy with motion toward the NGP being positive.

2 The line was drawn to separate the region containing most of the metal rich stars from the rest of the distribution. Although precise placement of the line’s intercept with the velocity axis does not have a significant effect on the calculated kinematics, the line was drawn low to minimize halo contamination of the DISK2 sample.

**Figure Captions**

Figure 1 Rotational velocity component (V) as a function of metallicity ([Fe/H]).

Figure 2 Total space velocity of stars in the sample plotted against height above the galactic plane (Z) in kiloparsecs. Points above the dotted line are the HALO2 sample and those below are the DISK2 sample. The symbols denote stars in different abundance ranges; solid circles are [Fe/H]$\geq$-1.0, crosses -1.6$\leq$[Fe/H]<-1.0, and open squares [Fe/H]<-1.6.

Figure 3 A histogram of V velocities (30 km/s bins) for the HALO2 sample

Figure 4 V velocity versus U and W velocity for the kinematically unbiased HALO1 sample with one sigma error bars in each coordinate. The dashed line (V=-220km/s) separates prograde from retrograde V velocities.

Figure 5 A histogram of metallicities ([Fe/H]) of stars in the DISK2 sample

Figure 6 Mean U, V, and W velocities plotted as functions of $M_v$(RR) for the HALO2 sample. U = filled in circles. V = crosses. W = open squares. Lines A, B, C, and D mark the values of $M_V$(RR) adopted by Layden (1994) & this work, Layden et al. (1996), Chaboyer et al. (1998) and Feast and Catchpole (1997) respectively.

Figure 7 U, V and W velocity dispersions plotted as functions of $M_v$(RR) for the HALO2 sample. U = filled in circles. V = crosses. W = open squares. Lines A, B, C, and D mark the values of $M_V$(RR) adopted by Layden (1994) & this work, Layden et al. (1996), Chaboyer et al. (1998) and Feast and Catchpole (1997) respectively

**Table Captions**

TABLE 1. Summary of proper motion data. (a) Quoted values probably underestimate the actual errors. (b) Individual errors were not quoted in the NPM. The error quoted is an RMS error. (c) The full space velocity error is the square root of the sum of the squares of the tangential and radial velocities.

TABLE 2a. Proper motion data used in our sample. The first column is the for the star number in our database. There are gaps in this sequence where stars have not been included in the sample for this paper. Proper motions in R.A. are given in seconds of time per century. Proper motions in declination are given in seconds of arc per century. In cases where there is more than one source listed, the proper motion is the mean of those from the sources listed weighted by the inverse variances.

TABLE 2b. Distances, radial velocities, and [Fe/H] for stars not in Layden (1994). The first column is the for the star number in our database. The distances are derived from photometry obtained from Kinman (1997). The distance errors are determined by a standard Monte-Carlo error simulation. The [Fe/H] values from "Preston" are computed using the Layden (1994) ΔS to [Fe/H] relation for ΔS values from Preston (1959)

TABLE 3. Results of kinematic analysis of our RR Lyrae samples. (a) <U>, <V>, and <W> are calculated in the frame of the solar system and not the LSR. Solar motion relative to the Local Standard of Rest is (U,V,W)=(-9,+12,+7) (Mihalas & Binney, 1981). This motion should be reflected in <U>, <V>, and <W> for the samples.

TABLE 4a. Comparison of various local halo samples.

TABLE 4b. Comparison of various thick disk samples. (a) The numbers given in the table are from an analysis performed on the Edvardsson et.al (1993) disk population with stars having ages greater than 9 Gyr being "Older."

TABLE 5. Comparison of metal weak thick disk samples. (a) MWTD kinematics as calculated by Morrison, Flynn, and Freeman (1990)

**TABLE 1.** Summary of proper motion data. (a) Quoted values probably underestimate the actual errors. (b) Individual errors were not quoted in the NPM. The error quoted is an RMS error. (c) The full space velocity error is the square root of the sum of the squares of the tangential and radial velocities.

| **Source for Proper Motions** | **Number of Stars** | **Avg Error $\mu_\alpha$ (arcseconds/ century)** | **Avg Error $\mu_\delta$ (arcseconds/ century)** | **Average Full Space Velocity Error (km/s) (c)** |
|---|---|---|---|---|
| All Sources | 130 | 0.301 | 0.295 | 29.1 |
| NPM | 39 | 0.500(b) | 0.500(b) | 47.2 |
| HIP | 5 | 0.294 | 0.328 | 24.7 |
| TAC | 4 | 0.211 | 0.228 | 12.6 |
| WMJ | 2 | 0.180(a) | 0.175(a) | 11.2(a) |
| **Averaged Proper Motions (All Sources)** | **80** | **0.214** | **0.197** | **20.9** |
| HIP + ACRS | 2 | 0.161 | 0.104 | 9.7 |
| NPM + HIP + PPM | 2 | 0.147 | 0.137 | 10.5 |
| NPM + HIP + ACRS | 2 | 0.124 | 0.065 | 9.9 |
| TAC + HIP | 4 | 0.200 | 0.178 | 16.4 |
| NPM + TAC + HIP | 20 | 0.143 | 0.131 | 14.6 |
| NPM + TAC | 23 | 0.241 | 0.247 | 22.1 |
| NPM + HIP | 26 | 0.264 | 0.229 | 26.6 |

**TABLE 2a.** Proper motion data used in our sample. (a) The first column is the for the star number in our database. There are gaps in this sequence where stars have not been included in the sample for this paper. (b) Proper motions in R.A. are given in seconds of time per century. Proper motions in declination are given in seconds of arc per century. (c) In cases where there is more than one source listed, the proper motion is the mean weighted by the inverse variances.

| (a) | Name | Galactic Latitude | $\mu_\alpha$ (b) sec/cent | err($\mu_\alpha$) sec/cent | $\mu_\delta$ (b) ″/cent | err($\mu_\delta$) ″/cent | Pmot Source (c) |
|---|---|---|---|---|---|---|---|
| 1 | RY PSC | -62.89 | 0.258 | 0.014 | -0.816 | 0.224 | TAC,NPM |
| 4 | SW AND | -33.08 | -0.032 | 0.019 | -2.284 | 0.257 | TAC,NPM |
| 5 | RX CET | -77.65 | -0.158 | 0.019 | -6.266 | 0.177 | HIP,NPM |
| 6 | DR AND | -28.57 | 0.238 | 0.040 | -1.370 | 0.500 | NPM |
| 7 | XX AND | -23.64 | 0.478 | 0.013 | -3.516 | 0.134 | TAC,NPM,HIP |
| 8 | RR CET | -59.89 | 0.065 | 0.011 | -4.483 | 0.186 | TAC,NPM |
| 9 | CI AND | -17.62 | -0.007 | 0.027 | -0.393 | 0.217 | HIP,NPM |
| 10 | RV CET | -64.40 | 0.181 | 0.010 | -2.057 | 0.123 | PPM,NPM,HIP |
| 11 | RZ CET | -60.34 | 0.157 | 0.012 | 0.041 | 0.189 | TAC,NPM,HIP |
| 12 | X ARI | -39.84 | 0.449 | 0.009 | -8.918 | 0.131 | TAC,NPM,HIP |
| 13 | SV ERI | -53.47 | 0.090 | 0.010 | -5.017 | 0.151 | PPM,NPM,HIP |
| 15 | AR PER | -2.27 | -0.057 | 0.013 | -0.962 | 0.104 | HIP,TAC |
| 16 | RX.ERI | -33.88 | -0.108 | 0.008 | -1.113 | 0.100 | ACR,NPM,HIP |
| 19 | TZ AUR | 20.91 | -0.067 | 0.031 | -0.986 | 0.251 | HIP,NPM |
| 21 | RR GEM | 19.52 | -0.027 | 0.013 | -0.240 | 0.240 | WMJ |
| 22 | TW LYN | 27.54 | 0.002 | 0.032 | 0.334 | 0.273 | HIP,NPM |
| 24 | AL CMI | 15.35 | -0.081 | 0.033 | -0.510 | 0.500 | NPM |
| 25 | SZ GEM | 22.09 | -0.070 | 0.011 | -2.904 | 0.143 | TAC,NPM,HIP |
| 26 | SS CNC | 26.28 | -0.056 | 0.036 | -1.720 | 0.500 | NPM |
| 27 | XX PUP | 8.72 | -0.132 | 0.014 | -0.210 | 0.213 | HIP |
| 28 | DD HYA | 19.30 | -0.022 | 0.025 | -0.854 | 0.300 | HIP,NPM |
| 29 | AS CNC | 31.23 | 0.204 | 0.036 | -0.820 | 0.500 | NPM |
| 30 | TT CNC | 28.38 | -0.289 | 0.018 | -3.117 | 0.212 | HIP,NPM |
| 31 | ET HYA | 18.31 | -0.018 | 0.015 | -0.991 | 0.220 | TAC,NPM,HIP |
| 32 | GO HYA | 30.32 | -0.014 | 0.033 | -0.980 | 0.500 | NPM |
| 33 | DG HYA | 24.95 | -0.112 | 0.017 | -1.528 | 0.275 | TAC,NPM |
| 34 | DH HYA | 22.95 | -0.159 | 0.033 | -0.670 | 0.500 | NPM |
| 35 | TT LYN | 41.65 | -0.840 | 0.010 | -4.168 | 0.084 | TAC,NPM,HIP |
| 36 | XX HYA | 21.35 | 0.128 | 0.034 | -2.920 | 0.500 | NPM |
| 37 | SZ HYA | 25.93 | -0.041 | 0.032 | -3.964 | 0.486 | HIP,NPM |
| 38 | AQ CNC | 38.10 | -0.175 | 0.012 | -3.562 | 0.186 | TAC,NPM |
| 39 | RW CNC | 43.53 | 0.025 | 0.020 | -3.427 | 0.168 | HIP,NPM |
| 40 | WW LEO | 38.45 | -0.002 | 0.033 | -2.630 | 0.500 | NPM |
| 41 | UU HYA | 38.18 | -0.115 | 0.016 | -1.375 | 0.238 | TAC,NPM |
| 42 | X LMI | 53.70 | 0.140 | 0.043 | -2.000 | 0.500 | NPM |
| 43 | RR LEO | 53.10 | -0.120 | 0.012 | -0.952 | 0.128 | TAC,NPM,HIP |
| 44 | WZ HYA | 34.40 | -0.019 | 0.012 | -1.518 | 0.141 | TAC,NPM,HIP |
| 45 | V LMI | 57.84 | 0.165 | 0.038 | -3.010 | 0.500 | NPM |
| 46 | RV SEX | 43.38 | -0.053 | 0.017 | 0.250 | 0.270 | TAC,NPM |
| 47 | SZ LEO | 57.83 | -0.108 | 0.033 | -2.540 | 0.500 | NPM |
| 48 | TV LEO | 49.06 | 0.056 | 0.023 | 0.345 | 0.354 | TAC,NPM |
| 49 | AN LEO | 60.72 | 0.018 | 0.033 | -3.050 | 0.500 | NPM |
| 50 | RX LEO | 70.51 | 0.028 | 0.037 | -2.660 | 0.500 | NPM |
| 51 | AE LEO | 68.19 | 0.165 | 0.034 | -1.250 | 0.500 | NPM |
| 52 | TU UMA | 71.87 | -0.578 | 0.020 | -5.261 | 0.257 | TAC,NPM |
| 53 | AX LEO | 66.30 | -0.146 | 0.027 | -2.404 | 0.347 | HIP,NPM |
| 54 | SS LEO | 57.06 | -0.168 | 0.012 | -2.875 | 0.186 | TAC,NPM |
| 55 | SU DRA | 48.27 | -0.801 | 0.015 | -7.730 | 0.092 | TAC,NPM,HIP |
| 56 | ST LEO | 66.15 | -0.066 | 0.017 | -3.754 | 0.191 | HIP,NPM |
| 57 | AA LEO | 66.10 | -0.017 | 0.033 | -3.340 | 0.500 | NPM |
| 58 | X CRT | 49.49 | -0.010 | 0.011 | -3.832 | 0.136 | TAC,NPM,HIP |
| 59 | UU VIR | 60.89 | -0.292 | 0.012 | -0.443 | 0.151 | HIP,ACR |
| 60 | AB UMA | 67.86 | -0.160 | 0.014 | -1.528 | 0.125 | TAC,NPM,HIP |
| 61 | SW DRA | 47.33 | -0.477 | 0.017 | -0.864 | 0.091 | TAC,NPM,HIP |
| 62 | UV VIR | 62.28 | -0.174 | 0.033 | -1.790 | 0.500 | NPM |
| 63 | UZ CVN | 75.94 | -0.051 | 0.030 | -2.985 | 0.383 | HIP,NPM |
| 64 | S COM | 85.84 | -0.136 | 0.021 | -1.706 | 0.198 | HIP,NPM |
| 65 | SV CVN | 79.40 | 0.009 | 0.041 | -2.540 | 0.500 | NPM |
| 66 | BQ VIR | 60.23 | -0.017 | 0.033 | -1.380 | 0.500 | NPM |
| 67 | SW CVN | 79.80 | -0.079 | 0.041 | -1.980 | 0.500 | NPM |
| 68 | Z CVN | 73.35 | -0.063 | 0.016 | -3.094 | 0.169 | TAC,NPM |
| 69 | AS VIR | 52.61 | 0.058 | 0.019 | -3.636 | 0.287 | TAC,NPM |
| 70 | AT VIR | 57.40 | -0.414 | 0.010 | -2.291 | 0.124 | TAC,NPM,HIP |
| 71 | RY COM | 85.06 | -0.043 | 0.036 | -1.770 | 0.500 | NPM |
| 72 | ST COM | 81.24 | -0.170 | 0.019 | -3.398 | 0.156 | HIP |
| 73 | AV VIR | 70.82 | 0.034 | 0.014 | -3.751 | 0.164 | TAC,NPM,HIP |
| 74 | RV UMA | 62.06 | -0.322 | 0.029 | -3.837 | 0.251 | TAC,NPM |
| 75 | RZ CVN | 77.15 | -0.429 | 0.017 | -0.047 | 0.152 | HIP,NPM |
| 76 | SS CVN | 72.63 | 0.059 | 0.012 | -4.363 | 0.160 | HIP,NPM |
| 78 | UY BOO | 68.81 | 0.011 | 0.009 | -5.368 | 0.029 | ACR,NPM,HIP |
| 79 | RU CVN | 74.51 | -0.231 | 0.039 | 0.210 | 0.500 | NPM |
| 80 | W CVN | 70.96 | -0.161 | 0.007 | -1.502 | 0.117 | TAC,NPM,HIP |
| 82 | ST VIR | 53.65 | -0.049 | 0.012 | -2.120 | 0.190 | TAC |
| 83 | SW BOO | 67.75 | -0.377 | 0.041 | 0.120 | 0.500 | NPM |
| 84 | AF VIR | 59.16 | -0.397 | 0.019 | -0.044 | 0.238 | HIP,NPM |
| 85 | RS BOO | 67.35 | 0.007 | 0.028 | -0.640 | 0.350 | TAC,NPM |
| 86 | SZ BOO | 65.50 | -0.057 | 0.037 | -0.850 | 0.500 | NPM |
| 87 | TW BOO | 62.85 | -0.024 | 0.012 | -5.533 | 0.156 | HIP,NPM |
| 89 | BT DRA | 51.21 | 0.030 | 0.021 | -3.255 | 0.174 | HIP,NPM |
| 91 | UU BOO | 58.01 | -0.023 | 0.029 | -4.225 | 0.334 | TAC,NPM |
| 92 | TV LIB | 39.67 | 0.003 | 0.033 | 1.030 | 0.500 | NPM |
| 93 | TV CRB | 56.51 | -0.020 | 0.021 | -0.579 | 0.292 | HIP,NPM |
| 94 | CS SER | 45.43 | 0.158 | 0.033 | -2.780 | 0.500 | NPM |
| 95 | VY SER | 44.10 | -0.699 | 0.008 | -1.171 | 0.108 | TAC,NPM,HIP |
| 96 | ST BOO | 55.21 | -0.128 | 0.009 | -1.317 | 0.135 | TAC,NPM,HIP |
| 97 | AR SER | 44.26 | -0.257 | 0.020 | 1.098 | 0.257 | HIP,NPM |
| 98 | VY LIB | 28.84 | 0.007 | 0.014 | -5.265 | 0.187 | TAC,NPM,HIP |
| 99 | AN SER | 45.23 | -0.016 | 0.018 | -0.685 | 0.205 | HIP,NPM |
| 100 | AT SER | 42.45 | -0.009 | 0.019 | -0.915 | 0.288 | HIP,NPM |
| 102 | AV SER | 36.83 | 0.006 | 0.012 | 0.166 | 0.186 | TAC,NPM |
| 103 | v445 OPH | 28.44 | -0.059 | 0.016 | 0.633 | 0.189 | HIP,TAC |
| 104 | v413 OPH | 25.97 | -0.074 | 0.033 | -1.620 | 0.500 | NPM |
| 106 | RW DRA | 40.60 | -0.028 | 0.062 | -0.810 | 0.500 | NPM |
| 107 | GY HER | 41.71 | 0.024 | 0.042 | 1.120 | 0.500 | NPM |
| 108 | VZ HER | 34.58 | -0.162 | 0.012 | -1.675 | 0.159 | HIP,NPM |
| 110 | DL HER | 26.59 | 0.078 | 0.034 | -0.120 | 0.500 | NPM |
| 111 | ST OPH | 16.64 | -0.006 | 0.011 | -0.080 | 0.110 | WMJ |
| 112 | TW HER | 24.80 | -0.003 | 0.017 | -0.532 | 0.224 | TAC,NPM |
| 113 | v455 OPH | 13.53 | -0.219 | 0.022 | -2.343 | 0.313 | HIP |
| 114 | BC DRA | 28.48 | -0.509 | 0.040 | 3.419 | 0.169 | HIP,NPM |
| 115 | IO LYR | 19.98 | -0.096 | 0.039 | 2.190 | 0.500 | NPM |
| 116 | AE DRA | 25.41 | -0.230 | 0.058 | 1.260 | 0.500 | NPM |
| 118 | CN LYR | 14.70 | -0.008 | 0.038 | -1.610 | 0.500 | NPM |
| 119 | RZ LYR | 15.81 | 0.079 | 0.039 | 1.990 | 0.500 | NPM |
| 120 | EZ LYR | 16.24 | -0.013 | 0.048 | 1.310 | 0.822 | HIP |
| 121 | XZ DRA | 22.50 | 0.072 | 0.040 | 0.564 | 0.263 | TAC,NPM |
| 122 | BK DRA | 22.10 | -0.268 | 0.017 | 2.997 | 0.138 | HIP |
| 123 | BN VUL | 3.41 | -0.342 | 0.013 | -3.420 | 0.190 | TAC |
| 124 | XZ CYG | 16.98 | 1.013 | 0.037 | -2.500 | 0.330 | TAC |
| 126 | v341 AQL | -22.04 | 0.197 | 0.012 | -2.630 | 0.200 | TAC |
| 127 | AA AQL | -24.99 | -0.036 | 0.016 | -1.253 | 0.251 | TAC,NPM |
| 129 | DX DEL | -18.84 | 0.098 | 0.008 | 0.795 | 0.086 | TAC,NPM,HIP |
| 130 | UY CYG | -9.63 | -0.023 | 0.011 | -1.740 | 0.056 | HIP,ACR |
| 131 | BT AQR | -30.61 | 0.006 | 0.015 | -0.802 | 0.238 | TAC,NPM |
| 132 | RV CAP | -35.54 | 0.136 | 0.015 | -10.614 | 0.175 | HIP,ACR |
| 133 | CP AQR | -31.34 | -0.064 | 0.015 | -1.900 | 0.238 | TAC,NPM |
| 134 | SW AQR | -31.33 | -0.286 | 0.015 | -5.911 | 0.182 | HIP,NPM |
| 135 | DM CYG | -12.41 | 0.104 | 0.039 | -0.720 | 0.500 | NPM |
| 136 | SX AQR | -34.01 | -0.276 | 0.014 | -4.709 | 0.209 | TAC,NPM |
| 137 | CG PEG | -20.76 | -0.012 | 0.013 | -0.552 | 0.145 | HIP,NPM |
| 138 | AV PEG | -24.05 | 0.079 | 0.009 | -0.896 | 0.110 | TAC,NPM,HIP |
| 139 | TZ AQR | -44.33 | 0.029 | 0.016 | -0.517 | 0.257 | TAC,NPM |
| 140 | VV PEG | -30.41 | -0.004 | 0.035 | -1.220 | 0.500 | NPM |
| 141 | CZ LAC | -4.60 | -0.049 | 0.032 | 0.099 | 0.283 | HIP,TAC |
| 142 | CQ LAC | -14.55 | 0.028 | 0.043 | -0.150 | 0.500 | NPM |
| 144 | BH PEG | -38.36 | -0.177 | 0.009 | -6.382 | 0.113 | TAC,NPM,HIP |
| 145 | BO AQR | -58.82 | -0.056 | 0.034 | -1.210 | 0.500 | NPM |
| 146 | DZ PEG | -41.45 | 0.116 | 0.034 | -2.490 | 0.500 | NPM |
| 147 | BR AQR | -65.24 | 0.034 | 0.033 | -0.010 | 0.500 | NPM |
| 148 | AT AND | -18.09 | -0.076 | 0.012 | -5.143 | 0.136 | HIP,TAC |

**Table 2b.** Distances, radial velocities, and [Fe/H] for stars not in Layden (1994). (a) The first column is the for the star number in our database. (b) The distances are derived from photometry obtained from Kinman (1997). The distance errors are determined by a standard Monte-Carlo error simulation. (c) The [Fe/H] values from "Preston" are computed using the Layden (1994) ΔS to [Fe/H] relation for ΔS values from Preston (1959).

| (a) | Name | Galactic Latitude | d (b) kpc | err(d) kpc | Vr km/s | err(Vr) km/s | Vr Source | [Fe/H] | [Fe/H] Source (c) |
|---|---|---|---|---|---|---|---|---|---|
| 27 | XX PUP | 8.72 | 1.20 | 0.03 | 386 | 7 | LHHKH | -1.50 | LHHKH |
| 35 | TT LYN | 41.65 | 0.65 | 0.01 | -67 | 1 | LHHKH | -1.76 | LHHKH |
| 72 | ST COM | 81.24 | 1.35 | 0.03 | -68 | 7 | LHHKH | -1.26 | LHHKH |
| 120 | EZ LYR | 16.24 | 1.35 | 0.03 | -60 | 23 | LHHKH | -1.56 | LHHKH |
| 123 | BN VUL | 3.41 | 0.61 | 0.01 | -235 | 4 | LHHKH | -1.52 | LHHKH |
| 130 | UY CYG | -9.63 | 0.98 | 0.02 | -2 | 6 | LHHKH | -1.03 | LHHKH |
| 141 | CZ LAC | -4.60 | 1.10 | 0.02 | -120 | 5 | HIC | -0.68 | Preston |
| 148 | AT AND | -18.09 | 0.77 | 0.02 | -252 | 5 | HIC | -0.98 | Preston |

**TABLE 3.** Results of kinematic analysis of our RR Lyrae samples. (a) <U>, <V>, and <W> are calculated in the frame of the solar system and not the LSR. Solar motion relative to the Local Standard of Rest is (U,V,W)=(-9,+12,+7) (Mihalas & Binney, 1981). This motion should be reflected in <U>, <V>, and <W> for the samples.

| Sample | Sample Size | <U>(a)<br>err(<U>) | <V>(a)<br>err(<V>) | <W>(a)<br>err(<W>) | σ(U)<br>err(σ(U)) | σ(V)<br>err(σ(V)) | σ(W)<br>err(σ(W)) | <[Fe/H]><br>err(<[Fe/H]>) | σ([Fe/H])<br>err(σ([Fe/H])) |
|---|---|---|---|---|---|---|---|---|---|
| HALO1<br>[Fe/H]<-1.3 | 81 | 8.<br>20. | -197.<br>12. | -8.<br>10. | 180.<br>14. | 111.<br>9. | 93.<br>7. | -1.68<br>0.03 | 0.30<br>0.02 |
| DISK1<br>[Fe/H]>-0.9 | 26 | 8.<br>11. | -41.<br>11. | -29.<br>6. | 55.<br>8. | 58.<br>8. | 31.<br>4. | -0.54<br>0.07 | 0.34<br>0.05 |
| HALO2<br>see Fig 2 | 84 | -1.<br>21. | -219.<br>10. | -5.<br>10. | 193.<br>15. | 91.<br>7. | 96.<br>7. | -1.59<br>0.04 | 0.35<br>0.03 |
| DISK2<br>see Fig 2 | 46 | 9.<br>8. | -47.<br>8. | -23.<br>6. | 56.<br>6. | 57.<br>6. | 40.<br>4. | -0.95<br>0.09 | 0.63<br>0.07 |
| DISK2A<br>[Fe/H]<-1.0 | 22 | 12.<br>14. | -59.<br>14. | -19.<br>11. | 64.<br>10. | 64.<br>10. | 52.<br>8. | -1.44<br>0.08 | 0.39<br>0.06 |
| DISK2B<br>[Fe/H]>-1.0 | 24 | 6.<br>11. | -35.<br>11. | -27.<br>6. | 54.<br>8. | 54.<br>8. | 31.<br>4. | -0.52<br>0.07 | 0.34<br>0.05 |

**TABLE 4a.** Comparison of various local halo samples.

| Sample | Number of Stars | <U><br>err(<U>) | <V><br>err(<V>) | <W><br>err(<W>) | σ(U)<br>err(σ(U)) | σ(V)<br>err(σ(V)) | σ(W)<br>err(σ(W)) |
|---|---|---|---|---|---|---|---|
| This Paper, HALO1<br>[Fe/H]<-1.3 | 81 | 8<br>20 | -197<br>12 | -8<br>10 | 180<br>14 | 111<br>9 | 93<br>7 |
| This Paper, HALO2<br>See Fig 2 | 84 | -1<br>21 | -219<br>10 | -5<br>10 | 193<br>15 | 91<br>7 | 96<br>7 |
| LHH (1996) Halo 3<br>RR Lyraes; V and [Fe/H] selected | 162 | 9<br>14 | -210<br>12 | -12<br>8 | 168<br>13 | 102<br>8 | 95<br>9 |
| Layden (1995) Halo<br>RR Lyraes; [Fe/H]<-1.3 | ~200 | | -202<br>13 | | 166<br>14 | 109<br>9 | 95<br>9 |
| Chiba & Yoshi (1998)<br>RR Lyraes & K Giants; [Fe/H] < -1.6 | 124 | 16<br>18 | -217<br>21 | -10<br>12 | 161<br>10 | 115<br>7 | 108<br>7 |
| Norris (1986) Halo<br>[Fe/H]<-1.2 | ~500 | | -183<br>10 | | 131<br>6 | 106<br>6 | 85<br>4 |
| Morrison et al. (1990)<br>K Giants; [Fe/H]<-1.6 w/o MWTD | | | -195<br>15 | | 133<br>8 | 98<br>13 | 94<br>6 |
| Beers & Sommer-Larsen (1995)<br>Dwarfs; [Fe/H] <-1.5 | 887 | | | | 153<br>10 | 93<br>18 | 107<br>7 |
| Carney et al. (1996) Low Halo<br>Subdwarfs; [m/H]≤-1.5 & Z<2kpc | 150 | -20<br>13 | -193<br>7 | -3<br>4 | 152<br>10 | 104<br>8 | 95<br>7 |
| Carney et al. (1996) Low Halo<br>Subdwarfs; e of orbit >0.85 & Z<2kpc | 97 | -32<br>19 | -208<br>6 | 0<br>5 | | | |

**TABLE 4b.** Comparison of various thick disk samples. (a) The numbers given in the table are from an analysis performed on the Edvardsson et.al (1993) disk population with stars having ages greater than 9 Gyr being "Older."

| **Sample** | **Number of Stars** | **<U> err(<U>)** | **<V> err(<V>)** | **<W> err(<W>)** | **σ(U) err(σ(U))** | **σ(V) err(σ(V))** | **σ(W) err(σ(W))** |
|---|---|---|---|---|---|---|---|
| This Paper, DISK1<br>[Fe/H] > -0.9 | 26 | 8<br>11 | -41<br>11 | -29<br>6 | 55<br>8 | 58<br>8 | 31<br>4 |
| This Paper, DISK2A<br>See Fig 2 w/ [Fe/H]<-1.0 | 22 | 12.<br>14. | -59.<br>14. | -19.<br>11. | 64.<br>10. | 64.<br>10. | 52.<br>8. |
| This Paper, DISK2B<br>See Fig 2 w/ [Fe/H]>-1.0 | 24 | 6<br>11 | -35<br>11 | -27<br>6 | 54<br>8 | 54<br>8 | 31<br>4 |
| LHH (1996) Disk 3<br>RR Lyraes; V and [Fe/H] selected | 51 | 6<br>8 | -45<br>9 | -16<br>6 | 52<br>8 | 48<br>8 | 29<br>5 |
| Layden (1995) Thick Disk<br>RR Lyraes; [Fe/H] > -0.5 | ~50 | | -22<br>9 | | 49<br>7 | 44<br>7 | 34<br>6 |
| Edvardsson et al. (1993)<br>Older Disk F Dwarfs; Age > 9 Gyr (a) | 58 | 22<br>8 | -38<br>6 | -5<br>5 | 59<br>6 | 48<br>4 | 38<br>4 |
| Beers & Sommer-Larsen (1995)<br>Dwarfs; -1.0<[Fe/H]<-0.6 & Z < 1kpc | 349 | | | | 63<br>7 | 42<br>4 | 38<br>4 |

**TABLE 5.** Comparison of metal weak thick disk samples. (a) MWTD kinematics as calculated by Morrison, Flynn, and Freeman (1990)

| **Sample** | **<U>**<br>**err(<U>)** | **<V>**<br>**err(<V>)** | **<W>**<br>**err(<W>)** | **σ(U)**<br>**err(σ(U))** | **σ(V)**<br>**err(σ(V))** | **σ(W)**<br>**err(σ(W))** |
|---|---|---|---|---|---|---|
| DISK2A | 12.<br>14. | -59.<br>14. | -19.<br>11. | 64.<br>10. | 64.<br>10. | 52.<br>8. |
| MWTD (a) | 25.<br>20. | -52.<br>14. | -10.<br>14. | 65.<br>18. | 24.<br>16. | 40.<br>13. |